\documentclass[fleqn,usenatbib]{mnras2}

\usepackage{newtxtext,newtxmath}

\usepackage[T1]{fontenc}
\usepackage{ulem}

\DeclareRobustCommand{\VAN}[3]{#2}
\let\VANthebibliography\thebibliography
\def\thebibliography{\DeclareRobustCommand{\VAN}[3]{##3}\VANthebibliography}

\usepackage{graphicx}	
\usepackage{amsmath}	

\title[X$\&$C asteroids]{Dark asteroids exhibiting intermediate characteristics between C and X types}

\author[S. Hasegawa et al.]{
Sunao Hasegawa,$^{1}$\thanks{E-mail: hasehase@isas.jaxa.jp}
Chrysa Avdellidou,$^{2}$
Micha\"{e}l Marsset,$^{3}$
Ullas Bhat,$^{2}$
Marco Delbo,$^{4,2}$
Daisuke Kuroda,$^{5}$
\newauthor
Moe Matsuoka,$^{6}$
Masateru Ishiguro,$^{7,8}$
Cristina A. Thomas,$^{9}$
Francesca E. DeMeo$^{10}$
and Pierre Vernazza$^{11}$
\\
$^{1}$Institute of Space and Astronautical Science, Japan Aerospace Exploration Agency, 3-1-1 Yoshinodai, Chuo-ku, Sagamihara, Kanagawa 252-5210, Japan\\
$^{2}$School of Physics and Astronomy, University of Leicester, University Road, Leicester LE1 7RH, UK\\
$^{3}$European Southern Observatory (ESO), Karl-Schwarzschild-Strasse 2, 85748 Garching bei M\"{u}nchen, Germany\\
$^{4}$Universit\'{e} C\^{o}te d'Azur, CNRS--Lagrange, Observatoire de la C\^{o}te d'Azur, CS 34229 -- F 06304 NICE Cedex 4, France\\
$^{5}$Japan Spaceguard Association, Bisei Spaceguard Center 1716-3 Okura, Bisei, Ibara, Okayama 714-1411, Japan\\
$^{6}$Geological Survey of Japan, National Institute of Advanced Industrial Science and Technology, Tsukuba 305-8567, Japan\\
$^{7}$Department of Physics and Astronomy, Seoul National University, Gwanak-gu, Seoul 08826, Republic of Korea\\
$^{8}$SNU Astronomy Research Center, Seoul National University, Gwanak-gu, Seoul 08826, Republic of Korea\\
$^{9}$Department of Astronomy and Planetary Science, Northern Arizona University, NAU Box 6010, Flagstaff, AZ 86011, USA\\
$^{10}$Department of Earth, Atmospheric and Planetary Sciences, MIT, 77 Massachusetts Avenue, Cambridge, MA 02139, USA\\
$^{11}$Aix Marseille Universit\'{e}, CNRS, CNES, Laboratoire d'Astrophysique de Marseille, Marseille, France
}

\date{Accepted XXX. Received YYY; in original form ZZZ}

\pubyear{20??}

\begin{document}
\label{firstpage}
\pagerange{\pageref{firstpage}--\pageref{lastpage}}
\maketitle

\begin{abstract}
Large main-belt asteroids with visible geometric albedos below 0.1 are predominantly classified within the C- and X-complex spectroscopic classes.
C-type and dark X-type asteroids typically exhibit flat to slightly negative and positive spectral slopes, respectively. 
They are further distinguished by the presence (for C-types) or absence (for dark X-types) of a shallow absorption feature near 1.0--1.3 \micron. 
We serendipitously discovered that the asteroids 1093 Freda and 1390 Abastumani display spectral characteristics intermediate between these two classes, combining a positive visible-to-near-infrared spectral slope with a shallow absorption band. 
A search in the literature reveals additional asteroids with similar properties. 
The existence of such objects, spanning a continuum of spectral shapes between C- and dark X-types, may point to a common genetic origin. 
Their spectral behavior could be explained by the presence of cronstedtite, an Fe-rich serpentine, on their surfaces.

\end{abstract}




\section{Introduction}
Planetesimals are the first sizeable solid objects that formed in the protoplanetary disc during the first million years of the solar system evolution. 
Moreover, these bodies accreted at different heliocentric distances, resulting in a large variety of compositions.  
The decay of short-lived radioactive isotopes, such as ${^{26}}{\rm Al}$ and${^{60}}{\rm Fe}$, likely caused part of this population to undergo differentiation, forming, according to the classical picture, an iron core, an olivine-rich mantle, and a crust. 
Subsequent dynamical processes displaced many planetesimals from their formation locations and redistributed them throughout the Solar System, including into the asteroid main belt \citep{Tsiganis2005,Levison2009,Walsh2011,Vokrouhlicky2016,Raymond2017a,Avdellidou2024}.

Modelling \citep{Morbidelli2009,Klahr2020,RibeirodeSousa2024} and observations \citep{Delbo2017,Delbo2019,Ferrone2023} have managed to show that the original sizes of planetesimals are typically 50--100 km in diameter.
However subsequent collisions over 4.5 Gyr broke several of the leftover planetesimals that survived the planetary formation era, forming families of asteroid fragments (\citealt{Nesvorny2015a,Nesvorny2024d} and references therein). 
Surviving planetesimals are hidden among the millions of collisional fragments in the main belt, while recent studies have proposed techniques to pinpoint the primordial planetesimals \citep{Delbo2017,Delbo2019,Ferrone2023}.

Leftover planetesimals (or hereafter large asteroids) should have undergone limited radial mixing after settling into their current orbits due to the negligible effect from non-gravitational forces (Yarkovsky effect, \citealt{Vokrouhlicky2015}) on such large bodies. 
The natural next step after identifying them is to study their physical properties in order to infer their composition which can be used as input to dynamical models and trace back their history until the early solar system times \citep{BourdelledeMicas2022}. 

Asteroid geometric visible albedo and reflectance spectra are the key properties to characterise asteroid composition and link them to meteorites 
(\citealt{DeMeo2022,Marsset2024,Marsset2026Inpress,Broz2024b,Broz2024a} and references therein).
\citet{BourdelledeMicas2022} performed the first spectroscopic investigation of large inner main belt asteroids that do not belong to known asteroid families. 
These authors report that $\sim$30 per cent of the observed population are dark C-complex, D- and T-type asteroids.
Current planetary formation models invoke formation of these bodies at heliocentric distances beyond the current main belt and perhaps even beyond Jupiter \citep{Kruijer2017,Kruijer2020,Anderson2025}.

In general, dark asteroids exhibiting featureless spectra with slopes ranging from flat to red are of particular interest because they may retain primitive, hydrated, and potentially organic-rich material \citep[e.g., ][]{Nakamura2023,Yokoyama2023,Zega2025,McCoy2025,Glavin2025,Barnes2025}.
These dark asteroids classify as C-complex and X-, Xc-, T-, and D-types within the Bus--DeMeo spectroscopic taxonomy \citep{DeMeo2009}. 
These asteroids dominate the mass budget of the main asteroid belt \citep[e.g., ][]{DeMeo2013,DeMeo2014b}.
It is considered that dark asteroids represent the parent bodies of CI and CM carbonaceous chondrites and interplanetary dust particles (IDPs) \citep[e.g., ][]{Burbine2008,Vernazza2015,Nakamura2023,Lauretta2024}.

We are currently expanding the study of large asteroids at other regions of the main belt. 
To date, we have reported several discoveries made as byproducts of our spectroscopic exploration of large asteroids \citep{Hasegawa2021b,Hasegawa2022a,Hasegawa2022b}. 
Here, we present the serendipitous discovery of asteroids possessing intermediate properties between C- and dark X-type asteroids.

\section{Spectroscopic observations and Data Analysis}
To characterise the dark population of large asteroids, we conducted new spectroscopic observations in the near-infrared (NIR) wavelength range of five objects that reside in the outer main belt beyond the 7:3 mean-motion resonance with Jupiter and in the Cybele region, in accordance with the methodology of past research \citep{Avdellidou2021,Avdellidou2025,Delbo2026} . 
The observations were carried out using the SpeX instrument \citep{Rayner2003} with the MORIS high-accuracy guider \citep{Gulbis2011} at the NASA Infrared Telescope Facility (IRTF). 
The SpeX instrument was set to PRISM mode with a slit size of 0.8\arcsec~$\times$ 15\arcsec, providing coverage across wavelengths from 0.7 to 2.5 \micron~and a spectral resolution of $\sim$200. 
For each target, a well-studied solar analogue star \citep{Marsset2020b} was obtained at a similar airmass to correct for the spectral slope. 
A neighborhood G2-type star within 250\arcsec~of the asteroid was observed for optimal telluric correction. 
The obtained spectral data were reduced using the IDL-based spectral reduction Spextool (v4.1, \citealt{Cushing2004}). 
A summary of the observation details is shown in Table~\ref{tab:Circumstances}.

\begin{table*}
	\centering
	\caption{Observational Circumstances. 
	$^{a}$Apparent V-band magnitude ($V$), heliocentric ($R_{\rm h}$) and geocentric distances ($\Delta$), and phase angle ($\alpha$).
	These values were obtained from the NASA/Jet Propulsion Laboratory Horizons System (https://ssd.jpl.nasa.gov/horizons/app.html).
        Text in the parentheses after asteroid name indicates the name of the observational run.
	}
	\label{tab:Circumstances}
	\begin{tabular}{llclllllc} 
		\hline
		Num&Name&Observation&$V^{a}$&$R_{\rm h}^{a}$&$\Delta^{a}$&$\alpha^{a}$&Solar\\
		      &           &Date             &[mag]  &[au]                   &[au]              & [$\degr$]&analogue\\
		\hline
		1390&Abastumani (sp293)&2022-05-01&14.1 &3.528 &2.526 &2.1&SA107--998\\
		1390&Abastumani (adv)&2022-08-01&15.4 &3.504 &3.360 &16.8&SA107--684, (local HD123385)\\
		780&Armenia (adv)&2023-10-18&15.0 &3.201 &3.282 &17.6&SA98--978\\
		491&Carina (adv)&2023-10-18&14.5 &2.978 &2.673 &19.4&Hyades 64\\
		1093&Freda (sp311)&2024-04-07&14.9 &3.651 &2.900 &11.7&SA102--1081,SA105--56, SA107--684, SA98--978\\
		1093&Freda (adv)&2025-05-31&12.7 &2.631 &1.646 &6.6&SA107--684, (local HD13784)\\
		713&Luscinia (adv)&2025-08-03&14.6 &3.304 &2.666 &15.3&SA110--361, (local HD152907)\\
		\hline
	\end{tabular}
\end{table*}

Next, we retrieved from the available literature data in visible (VIS) and NIR wavelength range (\citealt{Zellner1985,Bus2002a,Sergeyev2022}, and \citealt{GaiaCollaboration2023}) (see Table 2).
In addition, for each of the asteroids presented in this work we retrieved their diameter and geometric visible albedo from \textit{IRAS} \citep{Tedesco2002}, \textit{AKARI} \citep{Usui2011}, and \textit{WISE} \citep{Masiero2011} and estimated an average value for each parameter (see Table~\ref{tab:Classification}).

\begin{table*}
	\centering
	\caption{Physical properties of the 16 dark primitive asteroids presented in this work, and their classification in the Bus--Binzel  \citep{Bus2002b} and Bus--DeMeo \citep{DeMeo2009} taxonomic schemes, using different wavelength regions of their spectra.
	Asteroids 409, 76, 117, 209, 491, 713, 128, and 206 and 212, 268, 76, 117, 209, 128, and 206 were previously classified in \citet{Bus2002b} (only VIS) and \citet{DeMeo2009,Hasegawa2024} (VIS--NIR), respectively.
	Newly observed and unclassified asteroids were classified using their closest spectral match to the mean visible spectra listed in table III of the \citet{Bus2002b} and by a web tool available on the Asteroid spectrum classification using Bus--DeMeo taxonomy (http://smass.mit.edu/busdemeoclass.html), respectively.
        Number in the parentheses in `Region' column indicates the family to which it belongs.
        The values in `Band (PFA)' and `Band (MGM)' columns indicate the central wavelength of the absorption band of the asteroids obtained using PFA and MGM methods, respectively.
        Bold values in Band columns are those where the band centre value, when rounded, is more than 1.5 \micron.
        C04, D09, M16, D22, H22 and H24 in `NIR data' column stand for \citet{Clark2004,DeMeo2009,Marsset2016,DeMeo2022,Hasegawa2022b,Hasegawa2024}, respectively.
        }
	\label{tab:Classification}
	\begin{tabular}{lllcccccccl} 
		\hline
		Num&Name&Region&Diameter &$\mathrm{p_V}$&Bus--DeMeo&Bus--Binzel&Bus--DeMeo&Band&Band&NIR\\
		&&&[km]&&(VIS--NIR)&(only VIS)&(only NIR)&(PFA)&(MGM)&data\\
		&&&&&&&&[\micron]&[\micron]&\\
		\hline
                 203&Pompeja&Middle&112.7 &0.044 &X&Xk&X&---&---&H22\\
                 409&Aspasia&Middle&176.6&0.054&X&Xc&T&---&---&C04\\
                 212&Medea&Outer&144.9 &0.042 &X&X&X&---&---&H22\\
                 1390&Abastumani&Cybeles (1390)&98.6 &0.032 &X&X&C&\textbf{1.60}&\textbf{1.61}&This study (adv)\\
                  &&&&&&&&---&---&This study (sp293)\\
                 1093&Freda&Outer (31)&109.2 &0.045 &X&X&C&\textbf{1.46}&\textbf{1.53}&This study (adv)\\
                 &&&&&&&&---&\textbf{1.53}&This study (sp311)\\
                 268&Adorea&Outer (24)&138.6&0.045 &X&X&D&1.28&1.33&M16\\
                 76&Freia&Cybeles&165.8&0.046&X&X&D&1.33&1.31&D09\\
                 517&Edith&Outer&87.2&0.043&X&X&X&\textbf{1.55}&\textbf{1.52}&C04\\
                 536&Merapi&Cybeles&148.3 &0.047 &Xc&X&Cb&\textbf{1.64}&\textbf{1.69}&C04\\
                 117&Lomia&Outer&151.7 &0.051 &C&X&C&\textbf{1.48}&\textbf{1.53}&H24\\
                 780&Armenia&Outer (780)&99.8 &0.045 &C&X&C&1.31&1.35&This study (adv)\\
                 209&Dido&Outer&137.3 &0.050 &C&Xc&C&1.19&1.20&C04\\
                 &&&&&&&&1.08&1.06&H24\\
                 491&Carina&Outer&95.2 &0.068 &C&C&C&1.25&1.27&This study (adv)\\
                 713&Luscinia&Cybeles&100.3 &0.046 &C&C&C&1.29&1.41&This study (adv)\\
                 128&Nemesis&Middle (128)&176.2 &0.059 &C&C&Cb&1.25&1.29&H24\\
                 &&&&&&&&1.26&1.30&D09\\
                 &&&&&&&&1.19&1.21&D22\\
                 206&Hersilia&Middle&98.3&0.063 &C&C&Cb&1.26&1.33&H24\\
		\hline
	\end{tabular}
\end{table*}

\section{Result}
First, for each asteroid we combined VIS and NIR observations, where the spectral overlapping region shortward 1 \micron~between IRTF and VIS datasets verified the consistency between the observations. 
Next, we classified each individual VIS, NIR or combined VIS--NIR spectrum in the Bus--Binzel and Bus--DeMeo taxonomic schemes. 
Furthermore, to determine the absorption band position around 1.0--1.7 \micron, we employed two independent methods--the polynomial fitting approach (PFA) \citep[e.g., ][]{Gaffey1993b,Popescu2012} and the modified Gaussian model (MGM) \citep{Hiroi2000LPSC,Matsuoka2020}.
After removing the linear component of spectral slope, PFA used a sixth-order polynomial and MGM used a Gaussian function to determine the band position.

Results are reported in Table~\ref{tab:Classification}. 
Among the total sample, outer main belt asteroid 1093 Freda and Cybele asteroid 1390 Abastumani present distinct spectra.
Both asteroids are classified as X-type asteroids when using the VIS or VIS--NIR spectrum, but are classified as C-types when only the NIR part of the spectrum is considered. 
In addition, both asteroids exhibit an absorption feature around 1.5--1.6 \micron. 

   \begin{figure}
   \centering
   \includegraphics[width=8.5cm]{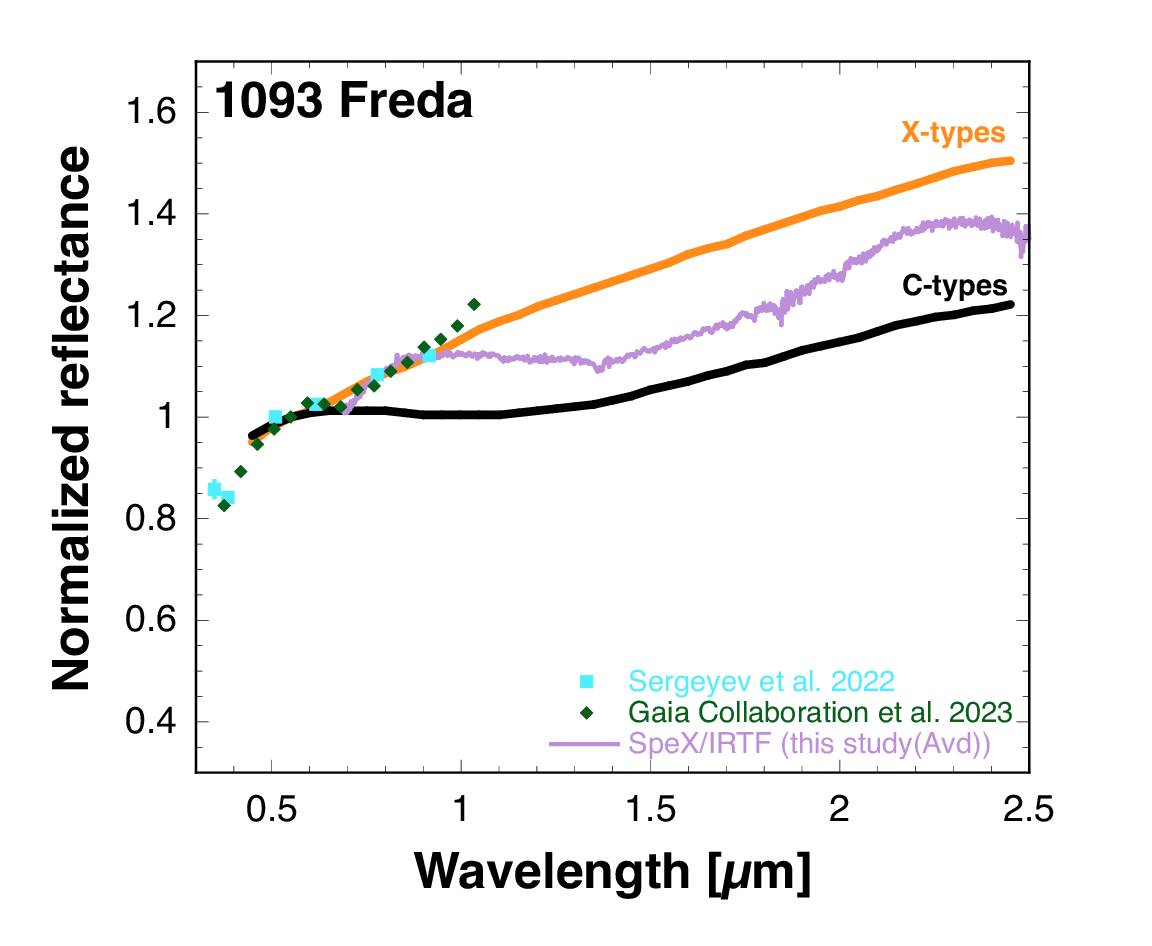}
   \includegraphics[width=8.5cm]{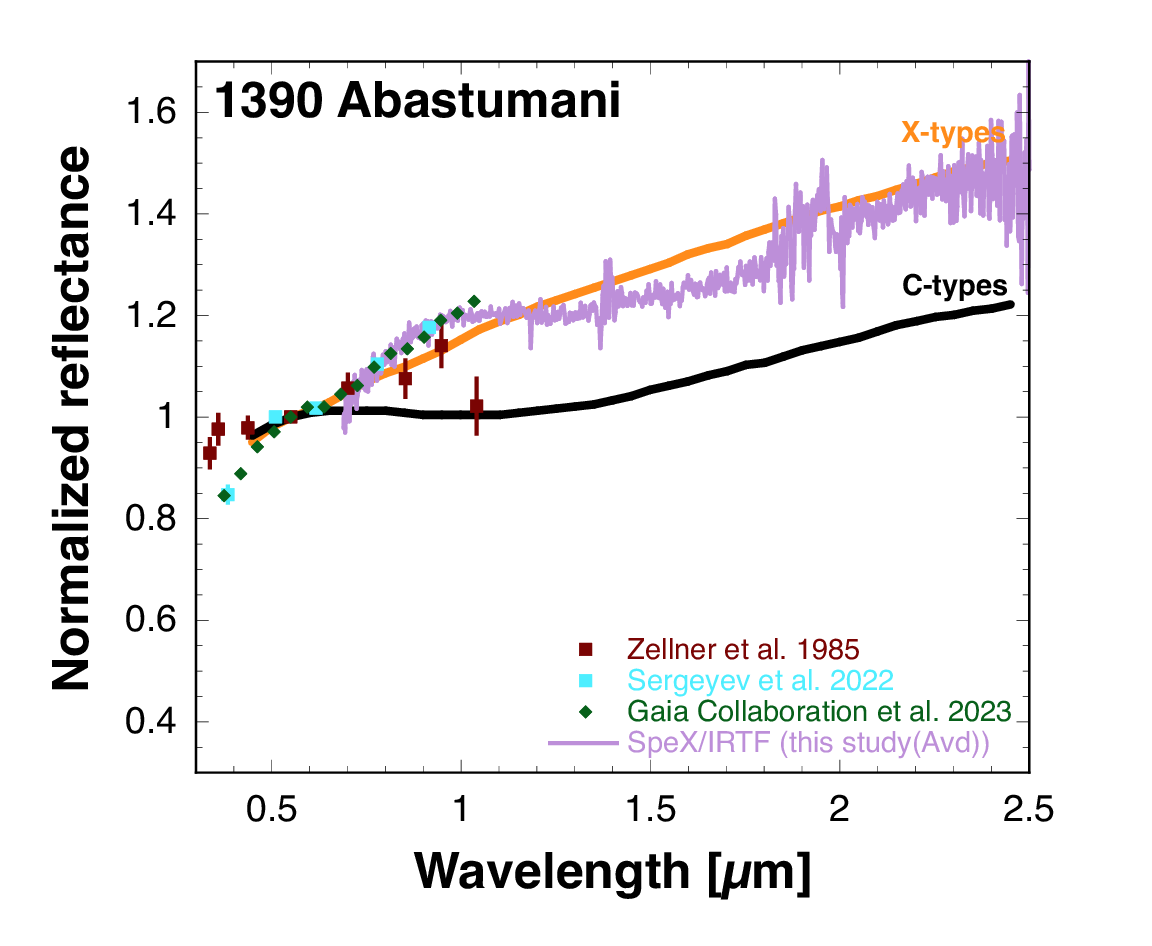}
      \caption{
      Reflectance spectra and photometric data for asteroids 1093 Freda and 1390 Abastumani. 
      The orange and black lines correspond to the average spectral values for X- and C-type asteroids from the Bus--DeMeo taxonomic scheme \citep{DeMeo2009}, respectively.
}
         \label{fig:1093-1390}
   \end{figure}

In general, X-type asteroids exhibit linearity with a moderate positive slope in the wavelength range from 0.45 to 2.45 \micron, but also shows sometimes a tendency for the slope to become less steep around 1.2 \micron~(Figure~\ref{fig:1093-1390}). 
However, unlike other X complex (Xe, Xk, and Xn), the X-type asteroids exhibit no absorption feature between 0.45 and 2.45 \micron. 
On the contrary, C-type asteroids, exhibit linear neutral or negative VIS spectral slopes, with positive slopes beyond 1.1--1.3 \micron~(Figure~\ref{fig:1093-1390}). 
This means that C-types have a shallow absorption feature around 1.1--1.3 \micron, while they may also present a slightly rough bump around 0.6 \micron. 
The presence of the feature around 1.5--1.6 \micron~and their reddish slope indicates that although these asteroids are classified as X-type asteroids, they also possess characteristics typical of C-type asteroids.

\section{Discussion}
\begin{figure*}
   \centering
   \includegraphics[width=4.41cm]{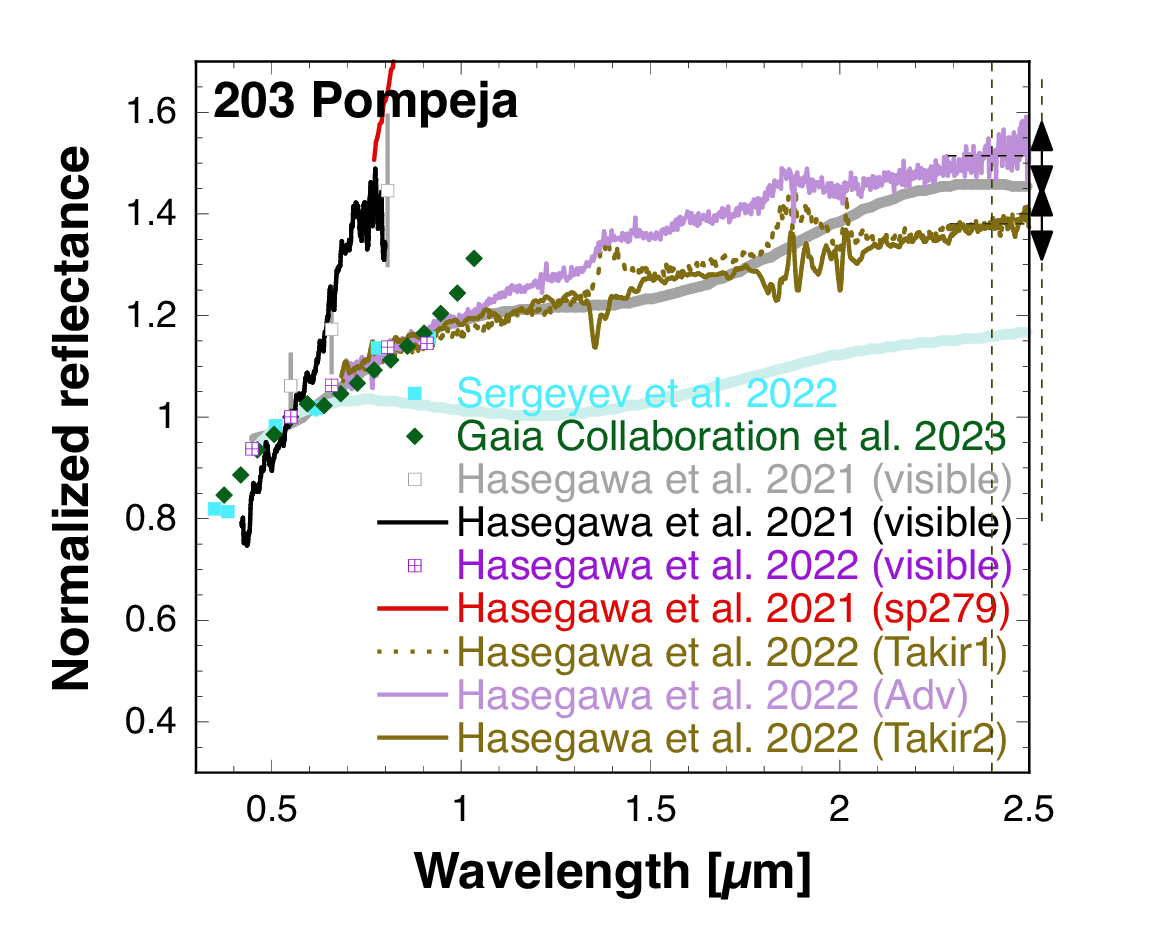}
   \includegraphics[width=4.41cm]{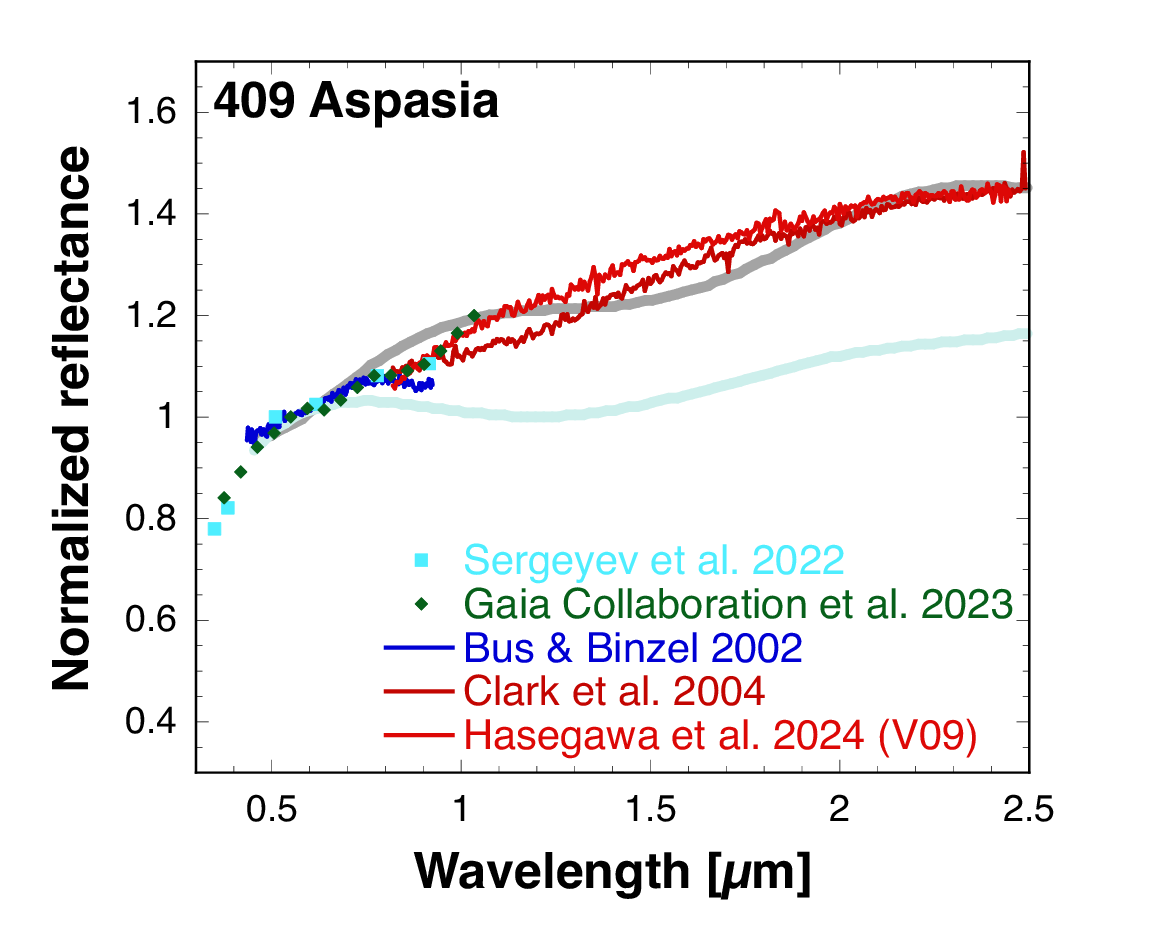}
   \includegraphics[width=4.41cm]{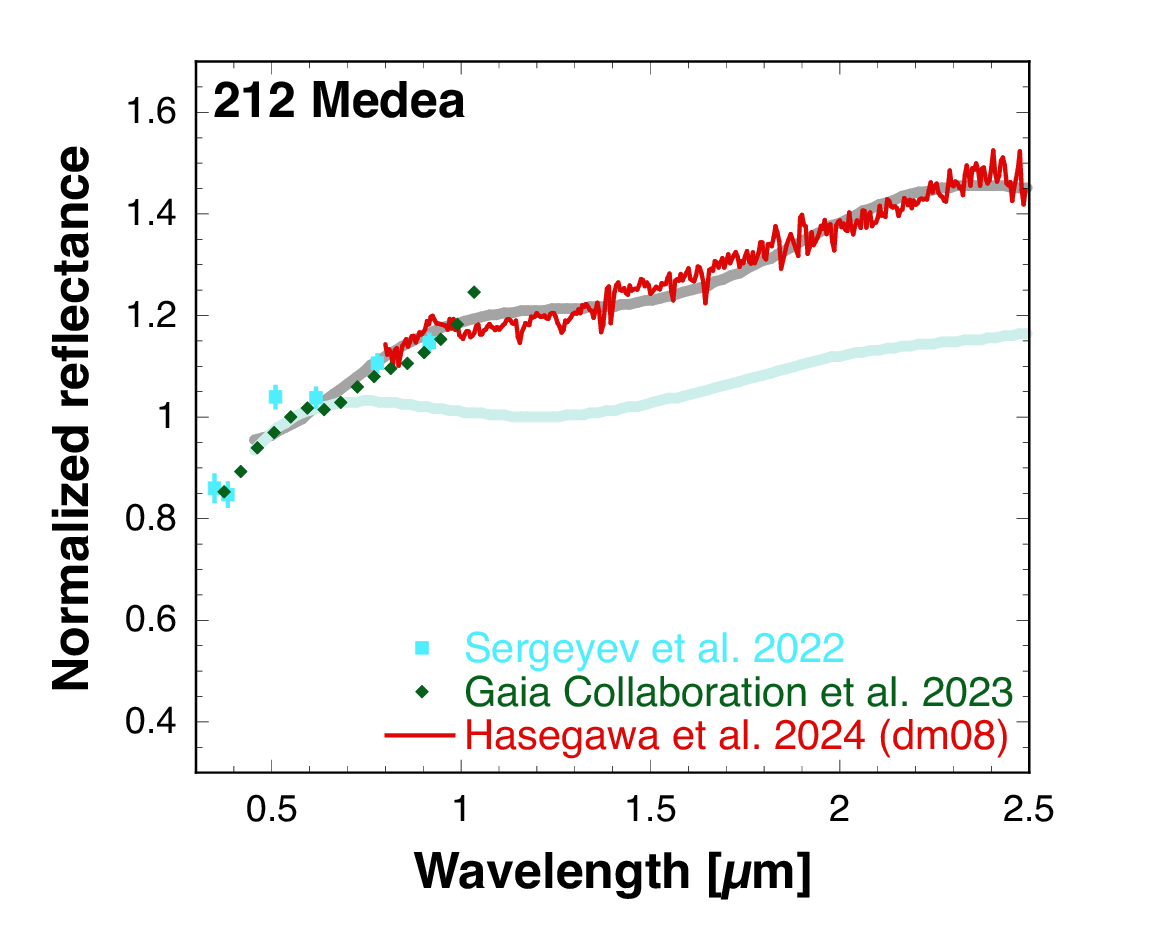}
   \includegraphics[width=4.41cm]{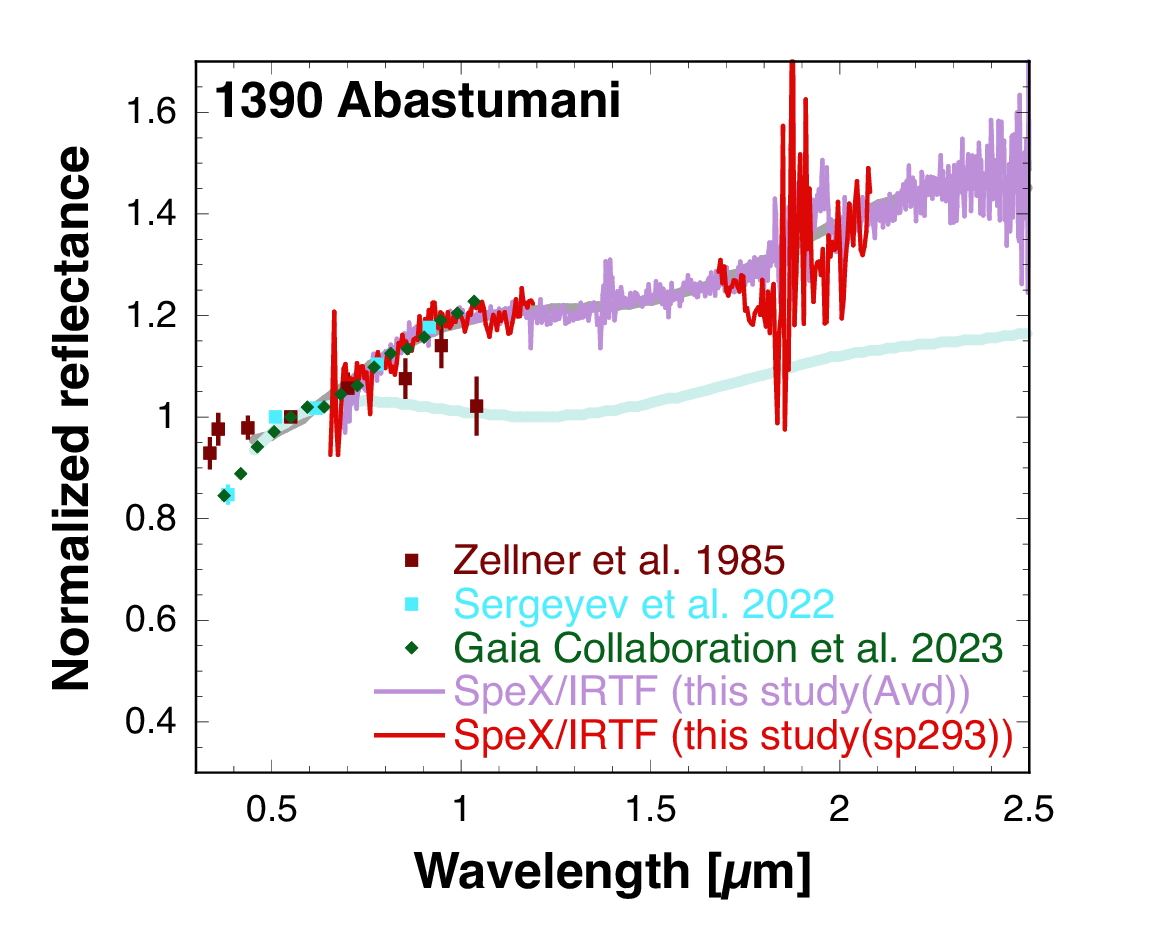}
   \includegraphics[width=4.41cm]{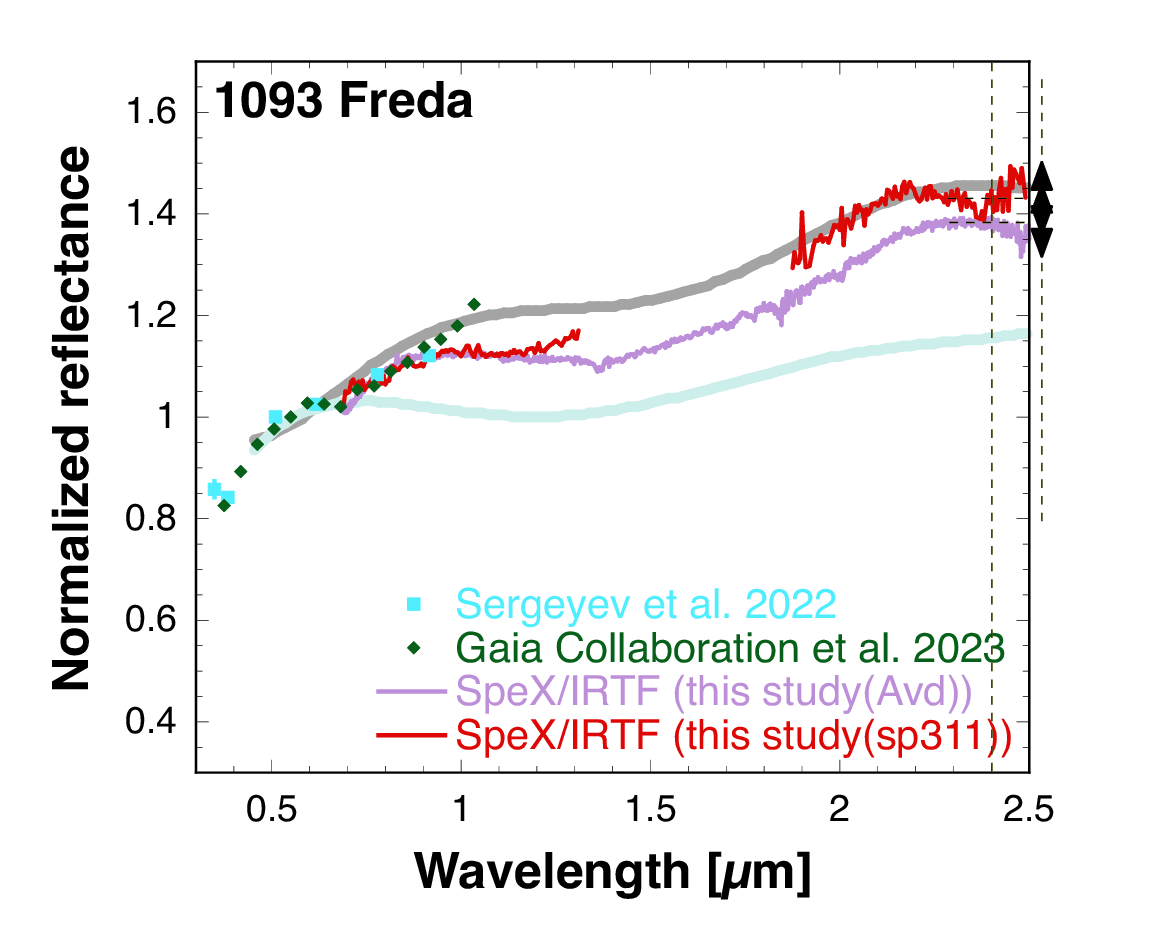}
   \includegraphics[width=4.41cm]{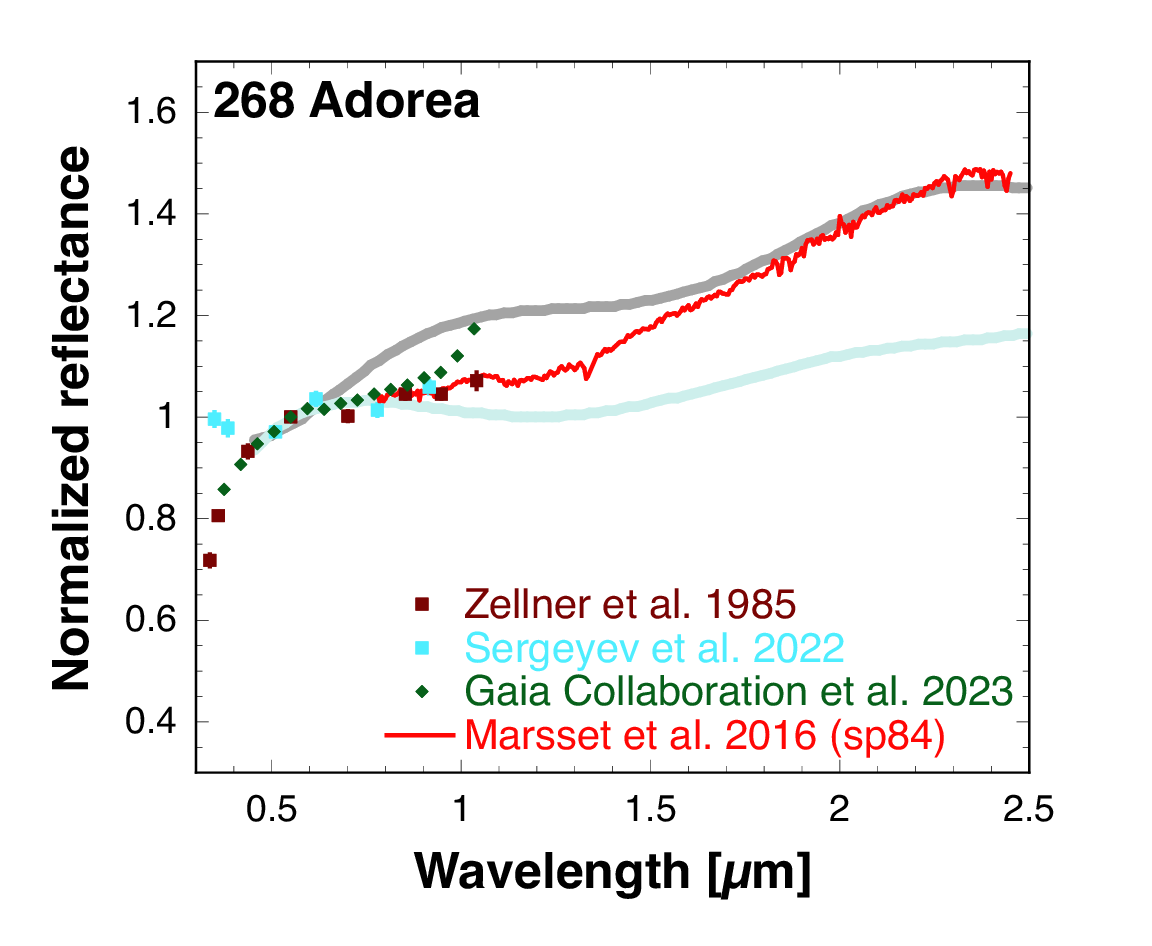}
   \includegraphics[width=4.41cm]{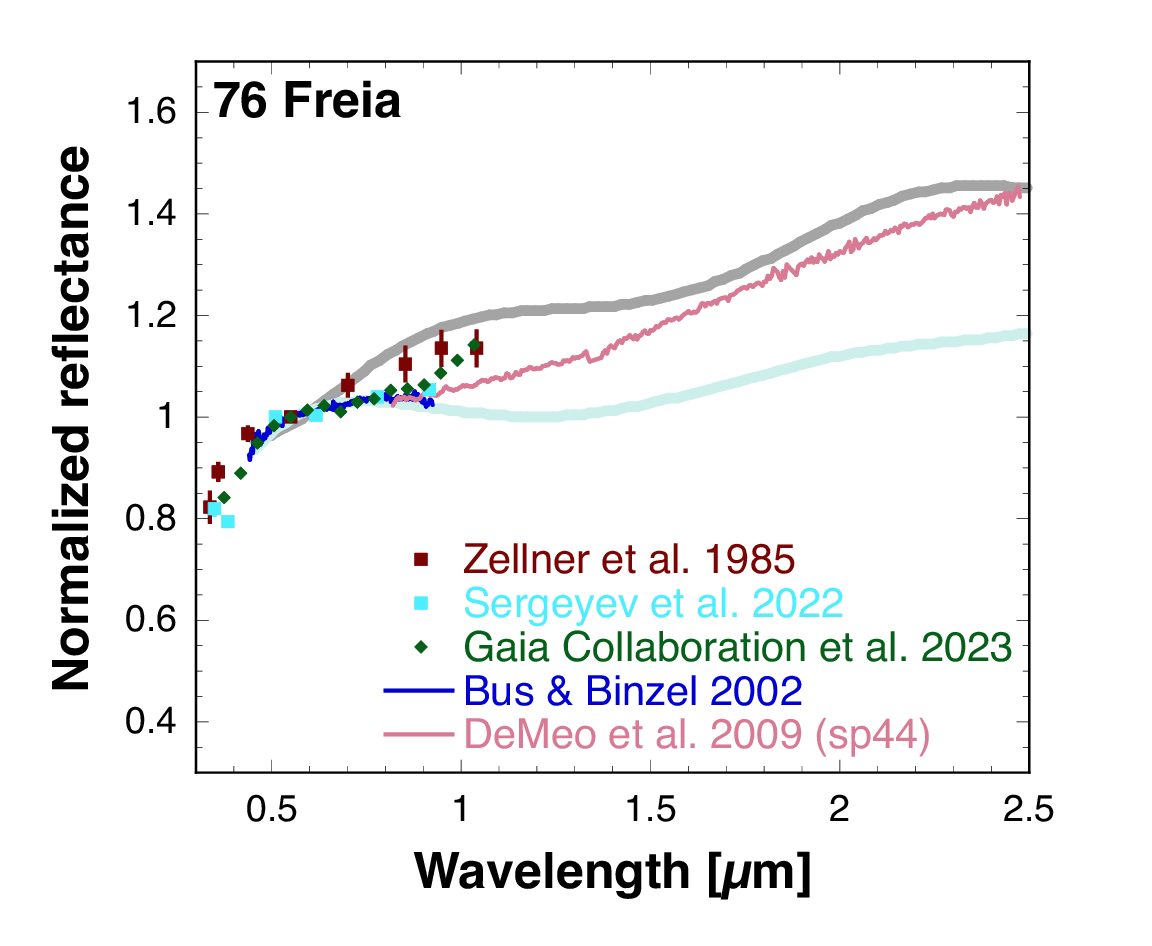}
   \includegraphics[width=4.41cm]{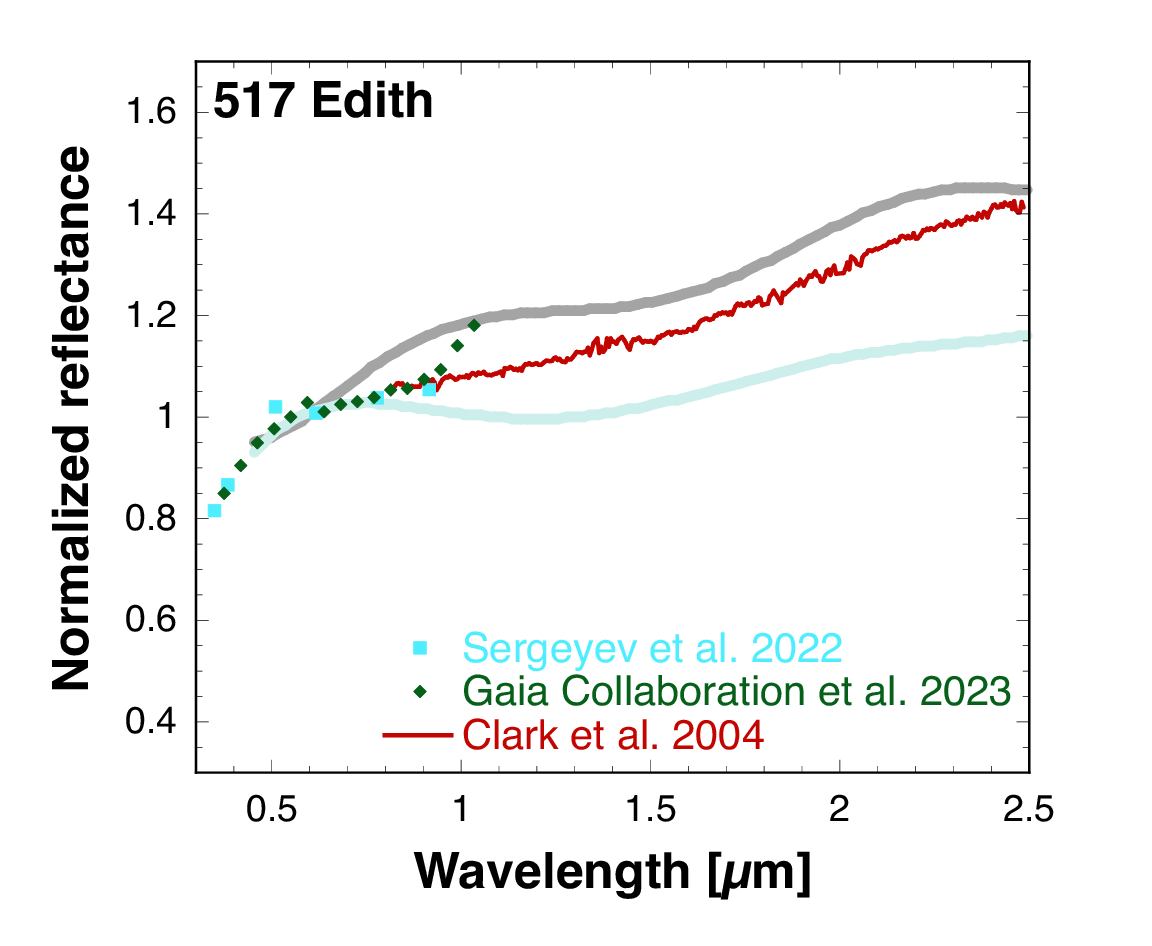}
   \includegraphics[width=4.41cm]{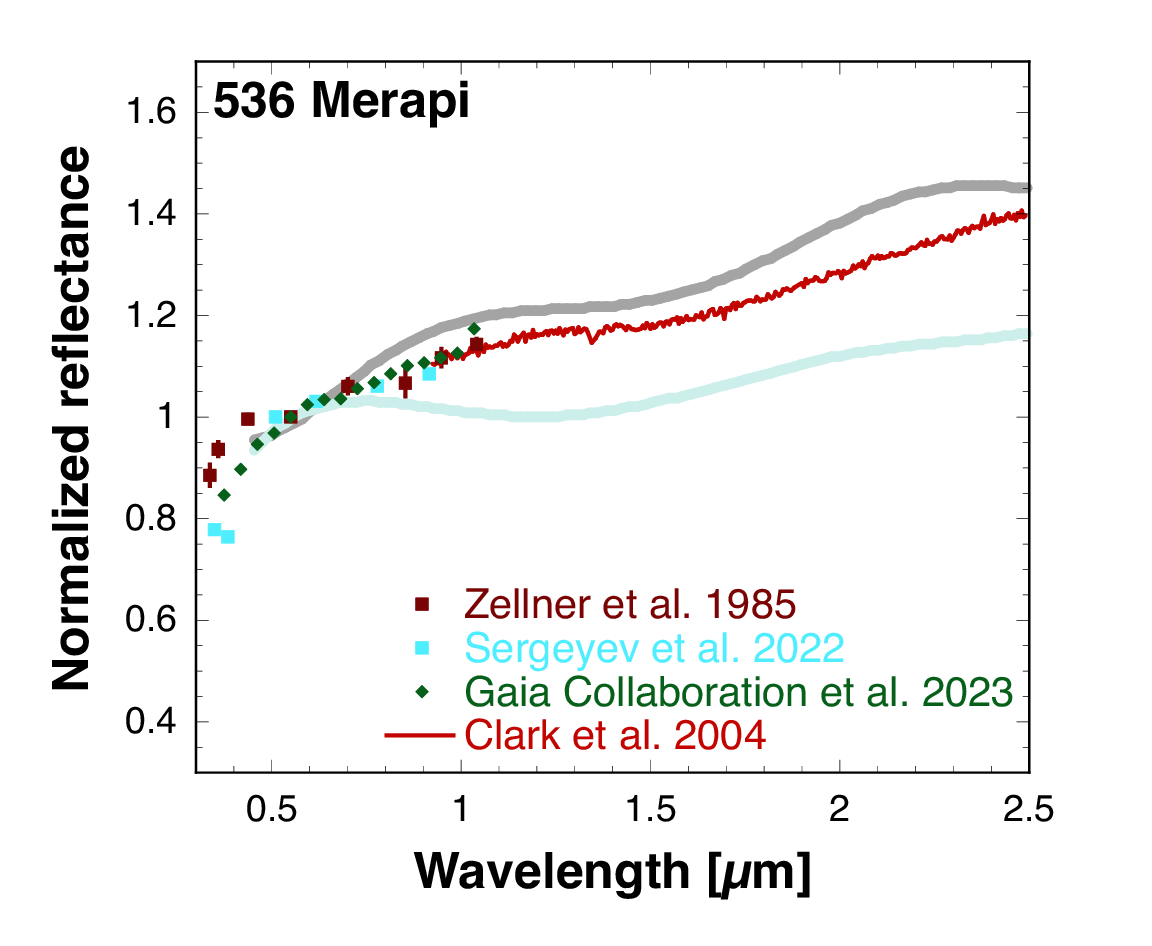}
   \includegraphics[width=4.41cm]{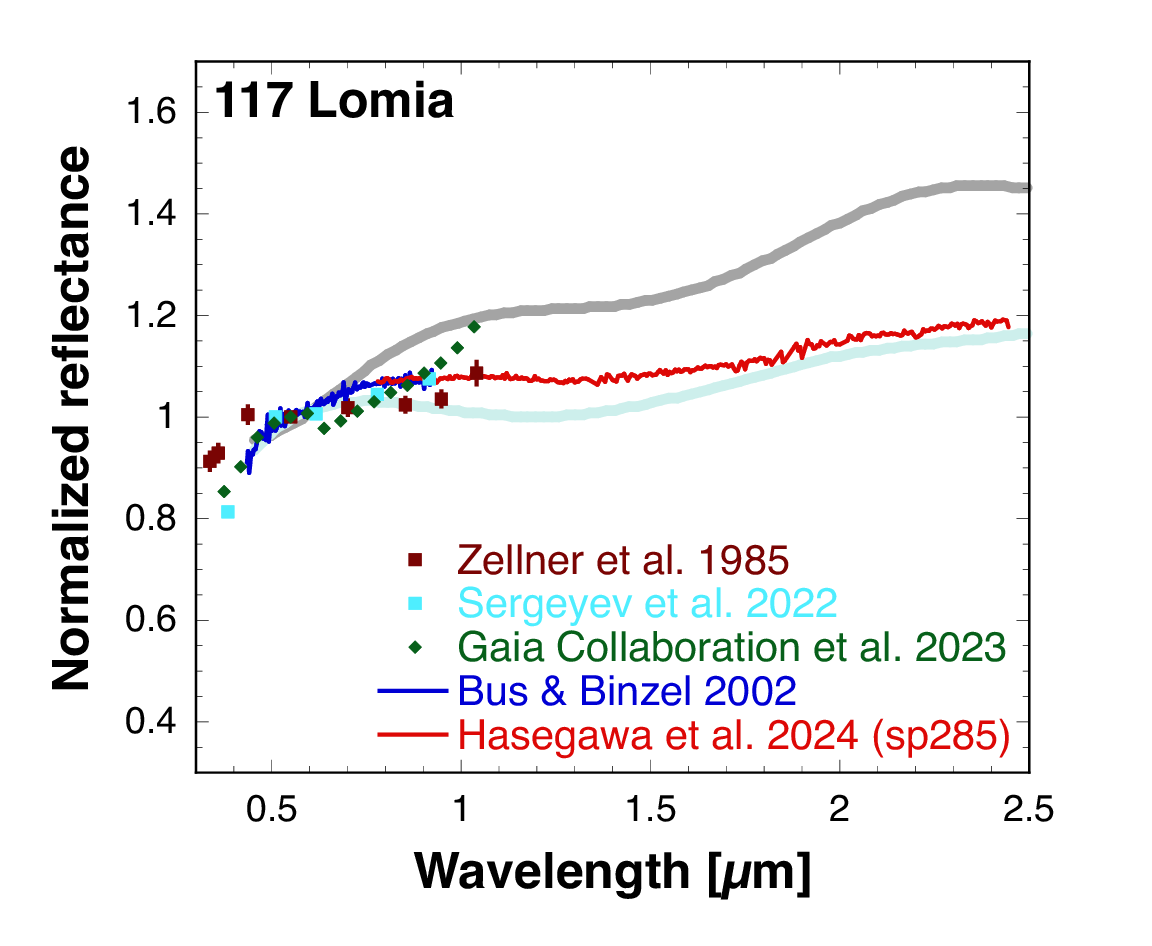}
   \includegraphics[width=4.41cm]{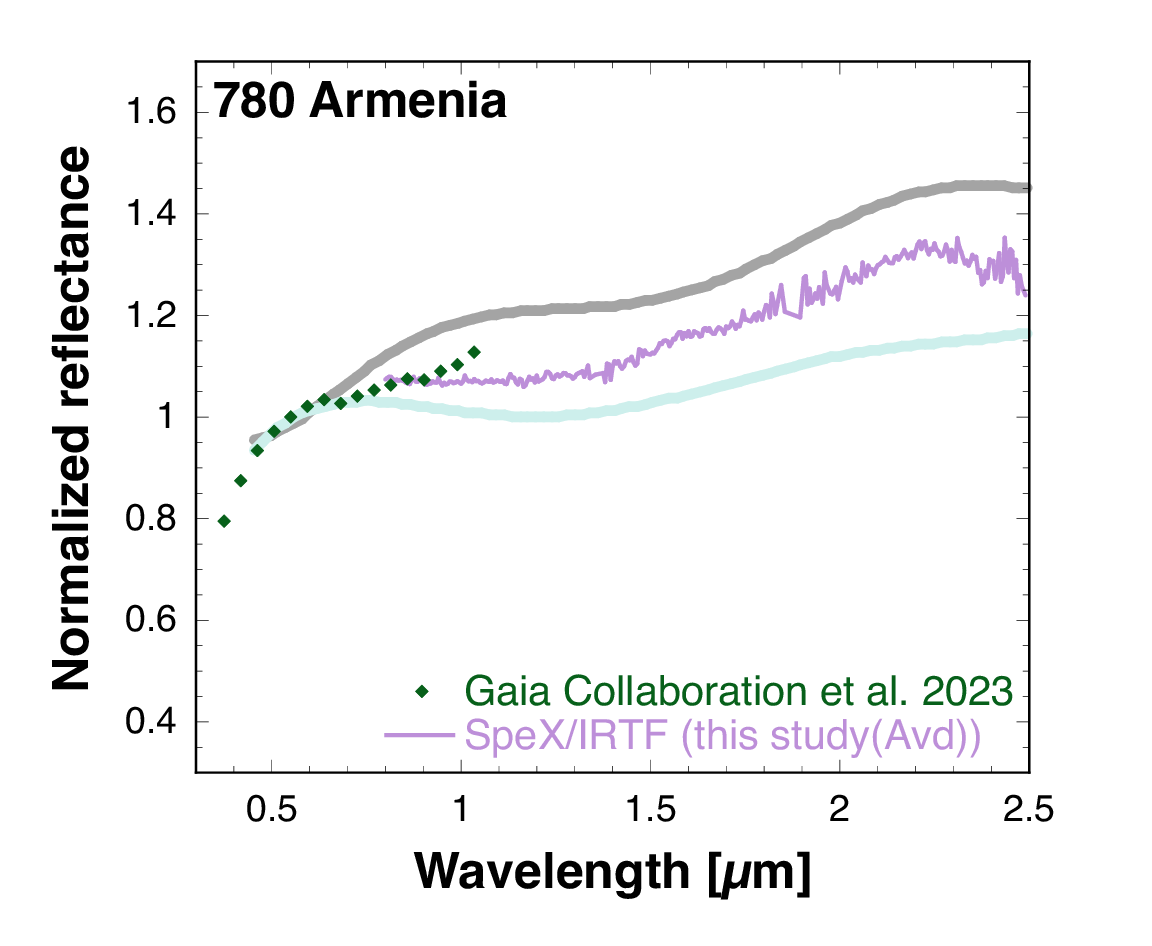}
   \includegraphics[width=4.41cm]{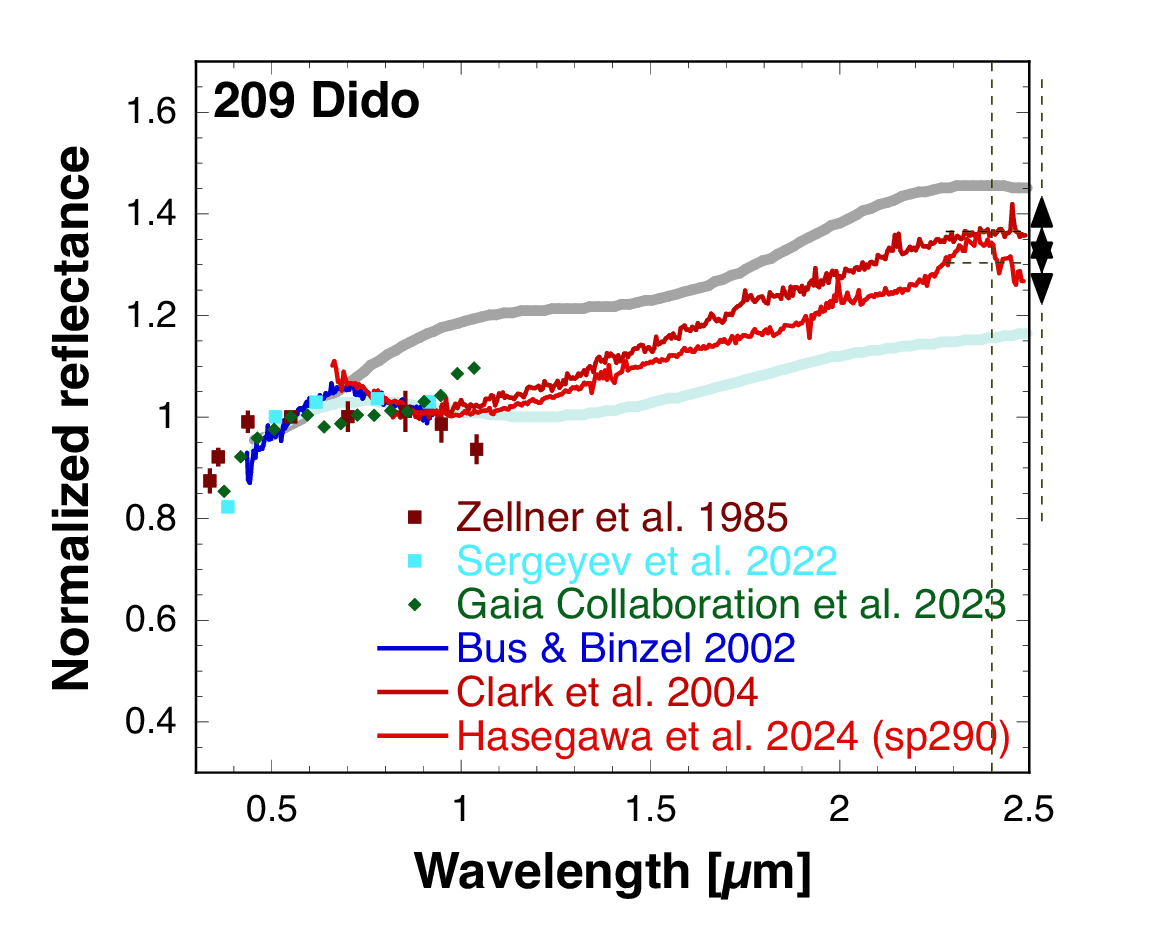}
   \includegraphics[width=4.41cm]{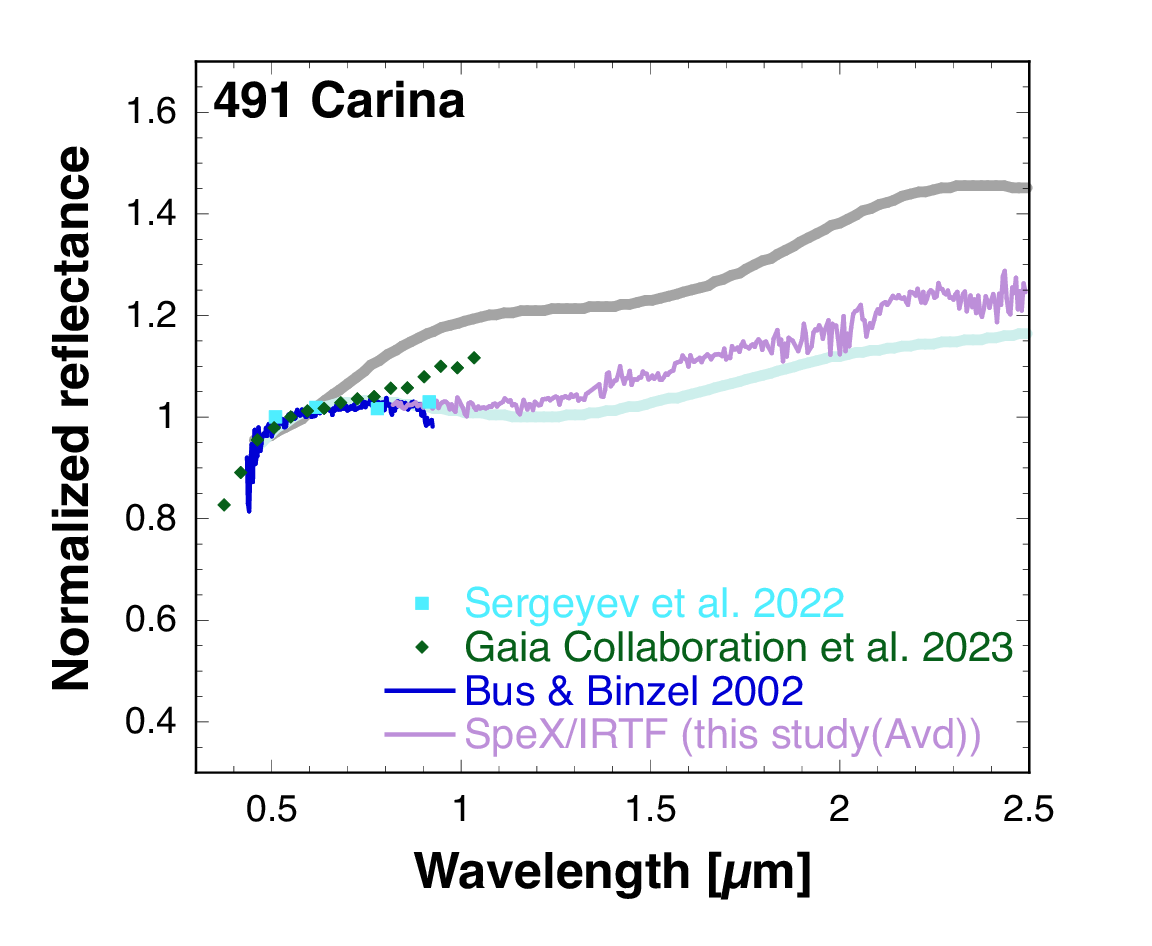}
   \includegraphics[width=4.41cm]{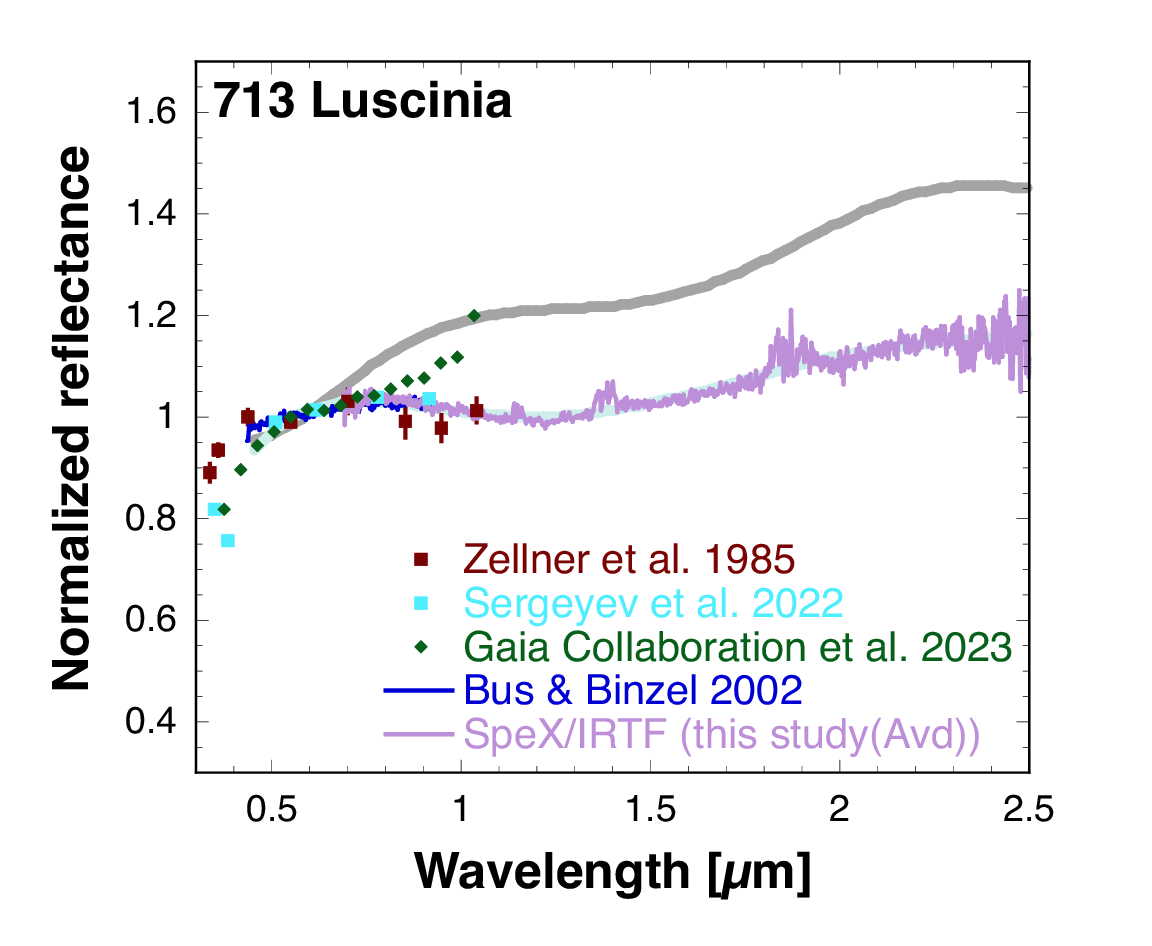}
   \includegraphics[width=4.41cm]{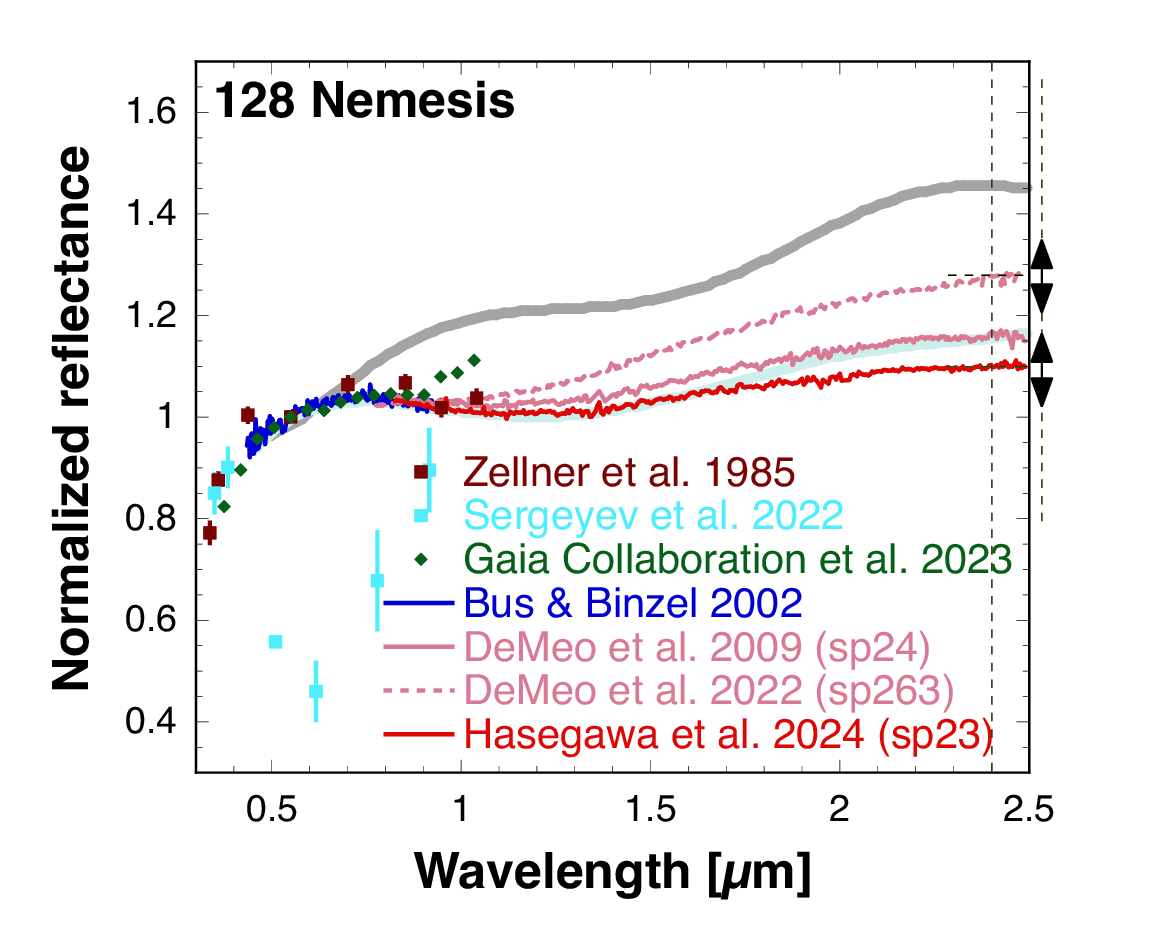}
   \includegraphics[width=4.41cm]{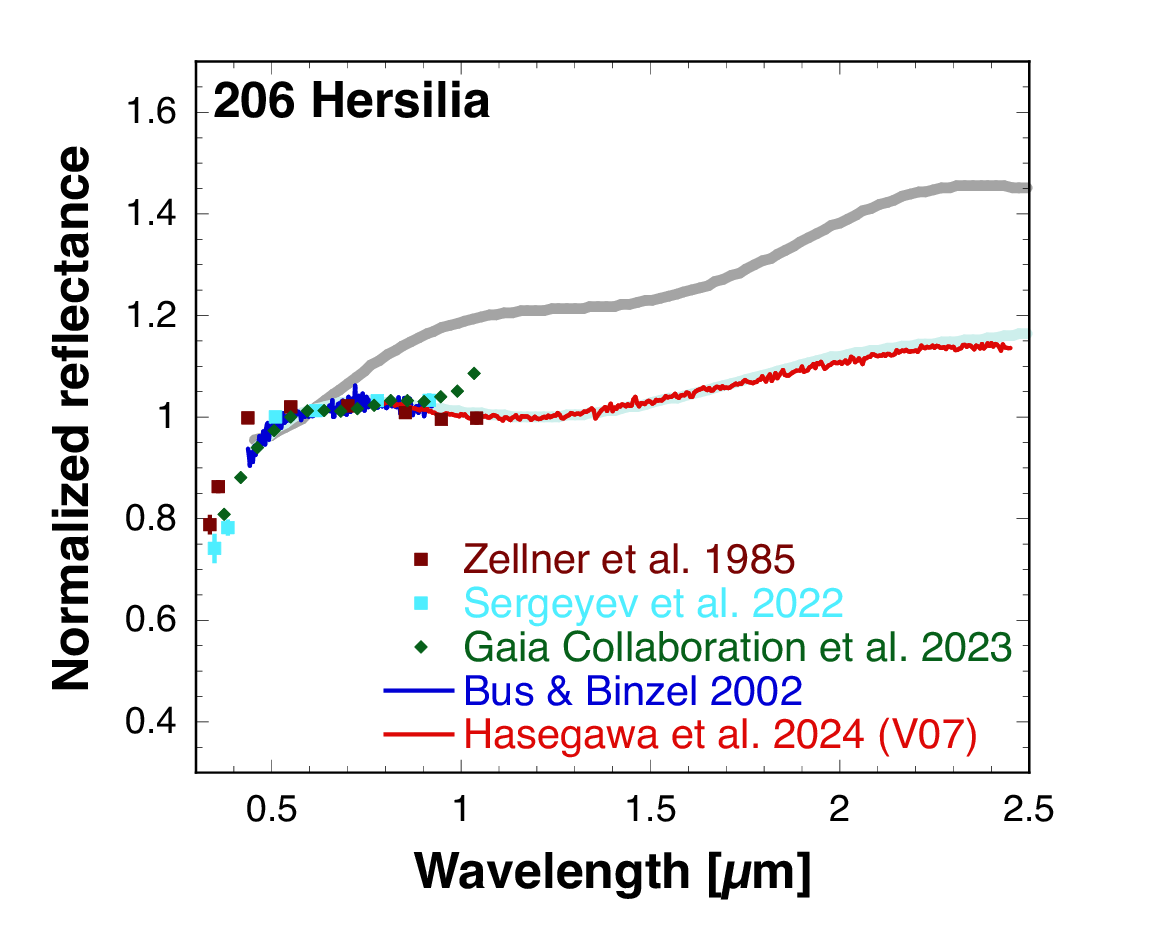}
      \caption{
VIS--NIR spectra and spectrophotometry for 16 large dark main belt asteroids. 
Spectra are normalised to 1 at 0.55 \micron. 
The black arrows show the slope uncertainty from \citet{Marsset2020b} centred at the reflectance value of the spectra at 2.40 \micron, as indicated by the black dashed lines. 
The gray and light blue spectral lines represent the reflectance spectra of 1390 Abastumani and 713 Luscinia, respectively.      
}
         \label{fig:XC}
   \end{figure*}

Asteroids 491 Carina, 713 Luscinia, 128 Nemesis, and 206 Hersilia are large dark C-type asteroids with reflectance spectra displaying a slight rough bump around 0.6 \micron~and a broad and shallow absorption feature around 1.1--1.3 \micron~on a slightly positive slope spanning 0.45--2.45 \micron~(Figure~\ref{fig:XC}). 
Further investigation of dark large asteroids with literature spectra shows that 268 Adorea, 76 Freia, 517 Edith, 536 Merapi, 117 Lomia, 780 Armenia, and 209 Dido show continuous spectral slopes between C-type asteroids like 713 Luscinia and X-type asteroids with C characteristics (hereafter X$\&$C asteroids) like 1390 Abastumani and the presence of the aforementioned shallow absorption feature (Figure~\ref{fig:XC}). 

It is known that absorption bands from olivine and pyroxene are present in the 1.0--1.5 \micron~region, which is the focus of this study \citep[e.g., ][]{Burbine2008,Cloutis2018}.
These minerals are major constituents of ordinary chondrites and aqueously unaltered carbonaceous chondrites (CO, CV, CR, and CK), but are not abundant in CI and CM chondrites \citep[e.g., ][]{Burbine2016,Cloutis2018,Cantillo2023}. 
The absorption features of these minerals are typically centred near $\sim$1.0--1.1 \micron~for olivine and $\sim$0.9 \micron~for pyroxene, which differs from the measurements obtained in this study.
Therefore, we do not think these minerals to be responsible for the spectral absorption we identified.

\subsection{The absorption feature around 1.0--1.3 \micron}
Magnetite is known as a mineral exhibiting absorption feature around 1.0--1.3 \micron~\citep{Cloutis2011a}. 
\citet{Izawa2019} demonstrated that the spectral positions of the absorption bands of magnetite samples for various grain sizes (ranging from less than 45 to 500--1000 \micron), porosities (fluffy, normal, packed dense, and pressed pallet), and highly substituted magnetite samples are located within the 1.0--1.3 \micron~range.
\citet{Chaves2023} showed that the band position of magnetite, simulated under micrometeoroid bombardment and solar wind irradiation, is located around the 1.0--1.3 \micron~range.
Spectral slope was also found to vary with the particle size of magnetite \citep{Izawa2019}. 
Although asteroid spectral slopes are known to depend on phase angle, this effect remains within the observational uncertainties reported by \citet{Marsset2020b} over the range of phase angles typical of main-belt observations, and can therefore be neglected  \citep{Hasegawa2022b}. 
The absorption band positions derived using the two methods employed in this study are consistent to within 0.15 \micron, which lies within the intrinsic width of the magnetite absorption feature.

Magnetite is present in CI1 chondrites \citep{Hyman1983}, the Tagish Lake and Tarda C2 unclassified carbonaceous chondrites \citep{Izawa2010,Marrocchi2021,Schrader2024}, and IDPs \citep{Christoffersen1986,Maupin2020}. 
It is also found on samples collected from 81P/Wild2, 162173 Ryugu, and 101955 Bennu \citep{Hicks2017, Nakamura2023, Lauretta2024}. 
Furthermore, \citet{Yang2010} indicated the presence of magnetite on B-type asteroids.
Note that the classification of asteroids in \citet{Yang2010} was based on the Bus--Binzel taxonomic scheme, which used only visible wavelengths.
When the Bus--DeMeo taxonomic scheme is used instead, the asteroids are classified as C or Cb type.
Combining these results, \citet{Yang2010} indicated that magnetite is present in asteroids classified as C- and Cb-types in the Bus--DeMeo taxonomy.
In addition, \citet{Carrozzo2026} demonstrated that the broad absorption feature around 1.2 \micron~observed on the surface of C-type object 1 Ceres originates from magnetite.
The presence of magnetite in C-type asteroids spanning diameters from $\sim$1000 down to $\sim$1 km suggests that magnetite is a widespread component of asteroids classified as C-types in the Bus--DeMeo taxonomic scheme.

On the other hand, the reflectance spectra of CI1 chondrites do not show absorption features around 1.0--1.3 \micron, as seen in C-type asteroids \citep[e.g., ][]{Johnson1973,Cloutis2011a}.
However, this absorption feature was found in the 162173 Ryugu returned samples, which was identified as having a very close composition to CI1 chondrites \citep{Amano2023}, and its laboratory spectrum closely resembles the spectrum obtained by telescopes \citep{Moskovitz2013,Perna2017,Marsset2022}.
Additionally, a broad and shallow absorption feature at 0.9--1.5 \micron~was observed in the 101955 Bennu returned sample \citep{Lauretta2024}.

\citet{Amano2023} demonstrated that spectra of CI1 chondrites underwent weathering in the terrestrial atmosphere, resulting in the adsorption of Earth's water, oxidation of iron, and formation of sulfates on the meteorite surface (as well as the interior because of the high porosity). 
As a result, it is thought that the shallow absorption around 1.0--1.3 \micron~originating from magnetite is hidden on terrestrial weathered CI1 chondrites. 
These results explain why the feature may be observed in fresh returned samples but not in terrestrially weathered CI1 chondrites.

This also suggests that CI1 chondrites that display not undergone terrestrial weathering should have a shallow absorption feature around 1.0--1.3 \micron~originating from magnetite, as shown in fresh returned samples.
In fact, an absorption band of spectrum of the newly discovered CI1 chondrite; Oued Chebeika 002 meteorite was observed around 1.3 \micron~
 \citep{Gattacceca2025}.

\subsection{The absorption feature around 1.5--1.6 \micron}
Here we discuss several possible origins for the 1.5--1.6 \micron~absorption feature observed in X$\&$C asteroids and assess their consistency with the spectral and meteoritic evidence.

\subsubsection{Maghemite}
X$\&$C asteroids such as 1093 Freda and 1390 Abastumani exhibit an absorption feature at slightly longer wavelengths than 1.3 \micron, which differs from the commonly known absorption position of magnetite \citep{Cloutis2011a,Izawa2019,Chaves2023}. 
However, maghemite, an iron oxide mineral similar to magnetite, is known as a mineral exhibiting absorption in this wavelength range (also see Figure~\ref{fig:magmag}).
Maghemite was found in comet 81P/Wild2 samples \citep{Hicks2017,Nguyen2017} as well as in the Orgueil CI1 carbonaceous chondrite \citep{Madsen1989,Gunnlaugsson1994}. 
However, maghemite is not present in the 162173 Ryugu returned sample and the Oued Chebeika 002 meteorite, which is a fresh CI1 carbonaceous chondrite \citep{Roskosz2024,Gattacceca2025}.
\citet{Gattacceca2025} showed that maghemite found in non-fresh CI1 carbonaceous chondrites such as Alais, Ivuna, and Orgueil are formed by terrestrial alteration.
Additionally, maghemite exhibits a sharp decrease in reflectance at wavelengths shorter than 0.75 \micron; however, this feature is not observed in these asteroids (Figure~\ref{fig:magmag}). 
Based on these two points, the absorption feature around 1.3--1.5 \micron~in X-type asteroids is unlikely to be caused by maghemite. 

\subsubsection{Regolith particle sizes}
\begin{figure}
   \centering
   \includegraphics[width=8.5cm]{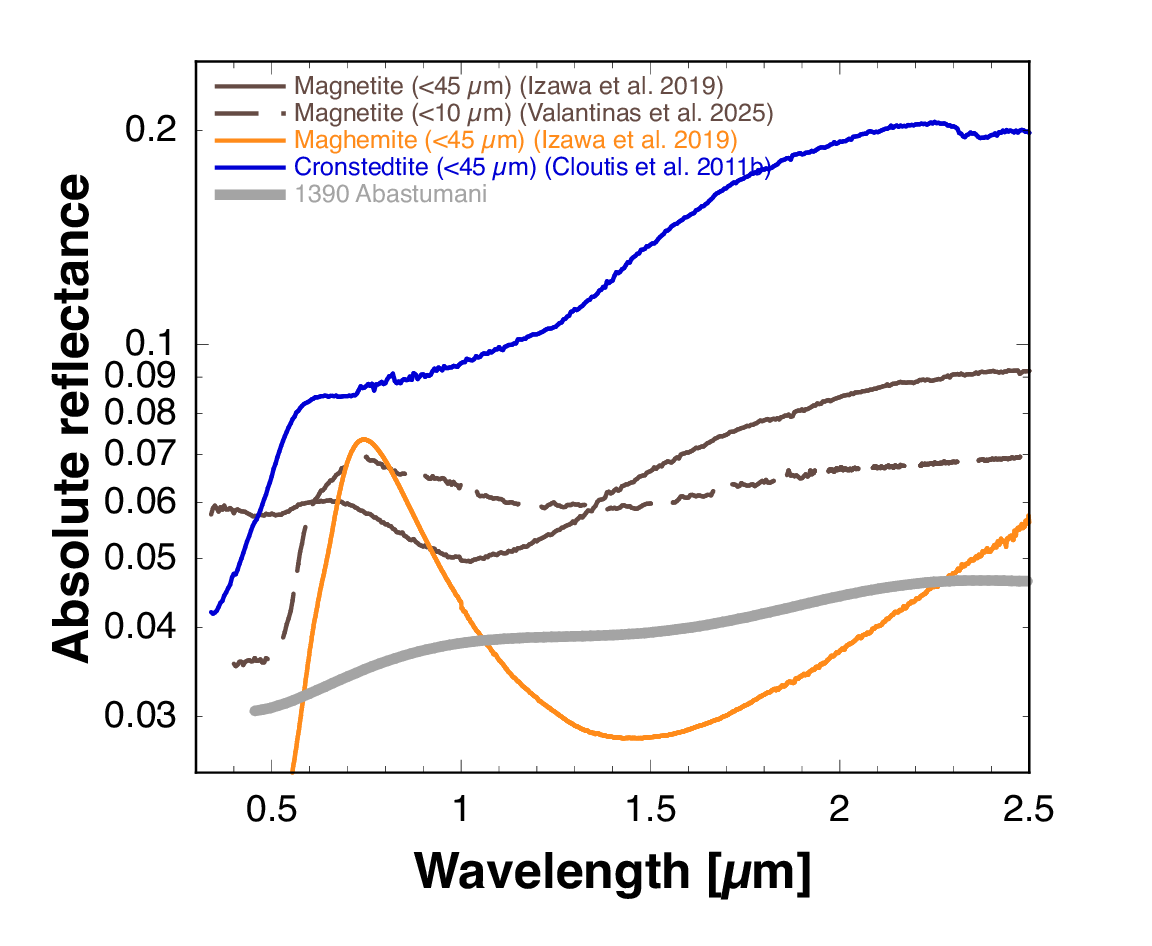}
      \caption{
The spectra of absolute reflectance of minerals and 1390 Abastumani.
Spectral data for minerals \citep{Cloutis2011b,Izawa2019}, with the exception of magnetite with particle sizes <10 \micron~\citep{Valantinas2025}, were obtained from the RELAB database \citep{Pieters2004LPSC}.
The absolute reflectance of 1390 Abastumani was derived from the spectrum obtained in this study and geometric albedo in Table~\ref{tab:Classification}.
}
         \label{fig:magmag}
   \end{figure}

When the particle size of magnetite becomes very fine (less than 10 \micron), the absorption position shifts towards longer wavelengths \citep{Valantinas2025}.
Indeed, \citet{Cantillo2023}, using variable particle diameters of Tarda C2 unclassified carbonaceous chondrite, observed the feature shifting towards longer wavelengths. 
Thus it is understood that as the particle diameter of meteorites becomes finer, the absorption position shifts towards longer wavelengths (Figure~\ref{fig:magmag}).
Moreover, it is known that the slope of the reflection spectrum becomes redder as the particle diameter becomes finer \citep[e.g., ][]{Johnson1973,Hasegawa2019}. 
The spectrum of the Tarda chondrite measured in \citet{Cantillo2023} similarly shows a redder spectral slope as particles become finer. 
These may imply that X$\&$C asteroids likely possess mineral compositions similar to other C-type asteroids, but simply exhibit finer particle sizes (less than 10 \micron) in their surface layers. 
However, \citet{Gundlach2013} showed that the regolith of large asteroids (D $\gtrsim$ 100 km) should be dominated by dust particles with sizes in the 10--100 \micron~range. 
This is incompatible with the much finer grain sizes (less than 10 \micron) required to reproduce the reported spectral properties of X$\&$C asteroids.
Note that fine particles of magnetite (less than 10 \micron) are embedded in the matrix of CI chondrites.
In this case, magnetite may remain on the asteroid surface even if the regolith particles themselves are larger than 10 \micron.
However, no absorption bands in other C-type asteroids have been observed around 1.5--1.6 \micron.
It is possible that these bands are obscured by other opaque minerals within the matrix and, as a result, remain undetected.

\subsubsection{Cronstedtite}
Cronstedtite, which is a Fe-rich serpentine, exhibits a broad, shallow absorption around 1.5 \micron~and a dip at 2.3--2.5 \micron~(Figure~\ref{fig:magmag}).
This shows that VIS and NIR spectra of X$\&$C asteroids could be  better explained by the presence of cronstedtite. 
However, cronstedtite has been found neither in CI1 chondrites nor in Ryugu samples \citep{Brearley2006,Ito2022}.
Cronstedtite is also absent in CM1 chondrites, which experienced more extensive aqueous alteration \citep{Suttle2021}, but is present in CM2 chondrites. 
Although the alteration sequences may differ between CI and CM parent bodies, if they are broadly comparable, the presence of cronstedtite may indicate a lower degree of aqueous alteration within the parent body.
Observational evidence for differentiation or partial differentiation has also been reported for CI-like parent bodies.
The outer main belt 31 Euphrosyne \citep{Yang2020c} and 24 Themis \citep{Marsset2016,Fornasier2016} families and the Jovian Trojan 3548 Eurybares family \citep{DeLuise2010} show an evolution from C- to X-type, consistent with parent body heterogeneity. 
In addition, within the Themis family ten asteroids show an absorption feature at 1.0--1.3 \micron, while three asteroids (461 Saskia, 468 Lina, and 621 Werdandi) exhibit absorption feature around 1.5--1.6 \micron~\citep{Marsset2016}. 
This indicates varying degrees of aqueous alteration, providing evidence that differentiation occurred within the parent body of Themis. 
128 Nemesis, the largest fragment of the Nemesis family, exhibits heterogeneity \citep{Hasegawa2024} showing varying absorption positions for different surface locations (Table~\ref{tab:Classification} and Figure~\ref{fig:XC}).
This may also indicate differences in hydration on the parent body. 
Heterogeneity is also detected among Jovian irregular satellites \citep{Vilas2024,Sharkey2025} that include C- and X-type asteroids. 
Four Jovian satellites (JX Lysithea, JVII Elara, JXII Ananke, and JVIII Pasiphae) exhibit absorption features around 1.5--1.6 \micron, whereas three Jovian satellites have no absorption features around 1.5--1.6 \micron~\citep{Sharkey2025}.
Differences in the occurrence of this absorption feature within the Jovian satellite system may suggest differentiation within this system. 
\citet{Vilas2024} also concluded that it is likely that the interior of the parent body of this Jovian system underwent aqueous alteration before it was disrupted.
Furthermore, among the four Jovian satellites that exhibit absorption around 1.5--1.6 \micron, the spectra of three of them in the 2.5--3.5 \micron~range are consistent with that of cronstedtite \citep[e.g., ][]{Bates2020}.
The above suggest that cronstedtite is present on these three satellites.


All the above observations together with our present work could provide input into the model of differentiation of CI chondrite parent bodies. 
According to \citet{Vernazza2017a,Vernazza2021}, the planetesimal core consists of CI1 chondrite material (C-types such as 491 Carina, 713 Luscinia, 128 Nemesis, and 206 Hersilia in Figure~\ref{fig:XC}), while the outer shell consists of volatile substances and chondritic porous IDPs (dark X-types such as 203 Pompeja, 409 Aspasia, and 212 Medea in Figure~\ref{fig:XC}). 
The current distribution of CI-like bodies and dark-type bodies in the asteroid belt supports this hypothesis \citep{Anderson2025}.

\begin{figure}
   \centering
   \includegraphics[width=8.5cm]{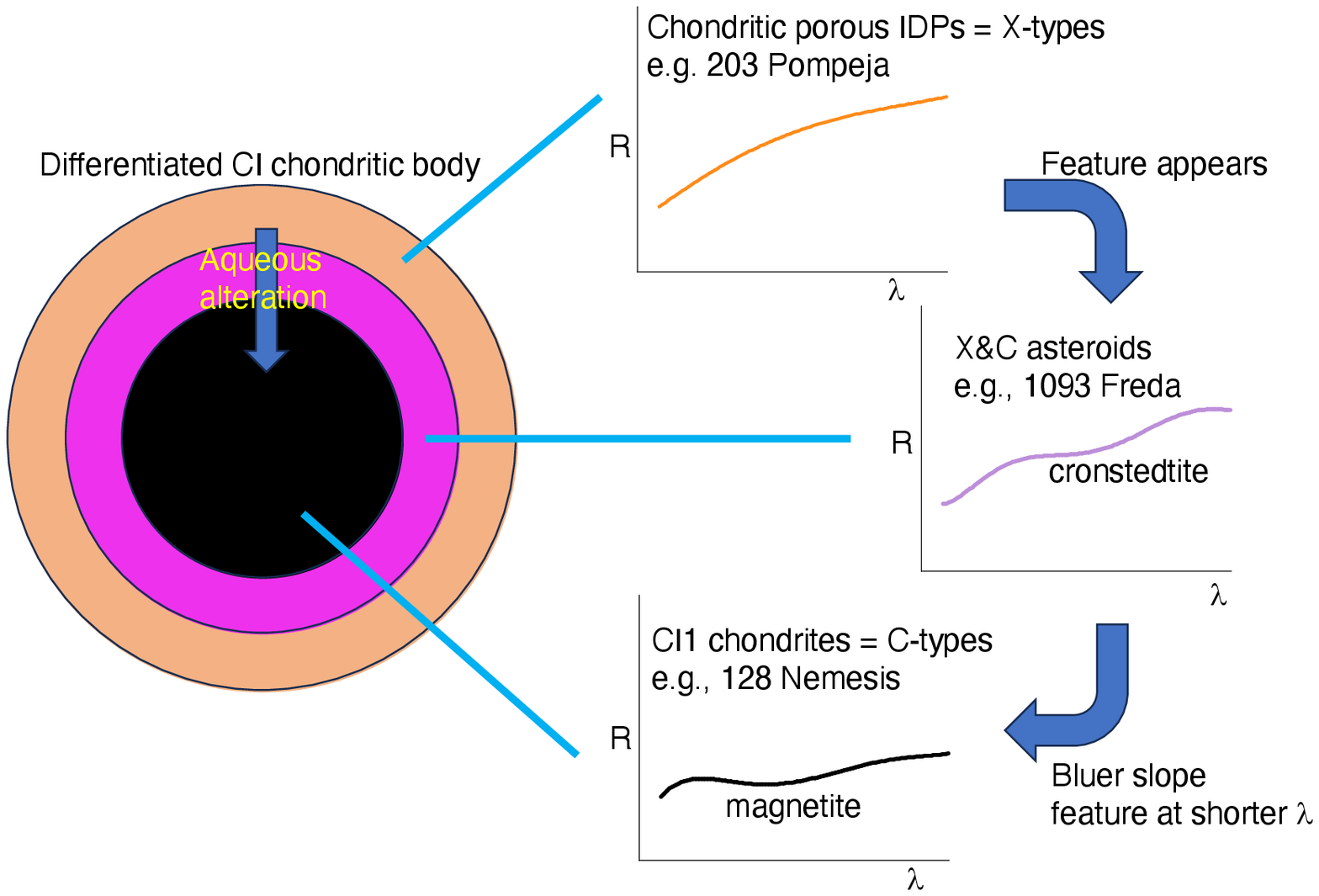}
      \caption{
      A summary schematic illustrating our proposed internal structure model for the partially differentiated parent bodies of CI chondrites and X$\&$C asteroids.
      This figure is adapted from \citet{Vernazza2017a,Vilas2024}.
}
         \label{fig:0DCI}
   \end{figure}

Considering the above, it is highly probable that  X$\&$C asteroids are immediate parent bodies of CI chondrites with a lower extent of aqueous alteration, formed in the mantle--the intermediate layer between the core and outer shell--of differentiated C-complex asteroids (Figure~\ref{fig:0DCI}).
Indeed asteroids 780 Armenia and 1390 Abastumani are parents of their respective families, while asteroids 268 Adorea and 1093 Freda are very large family members within Themis and Euphrosyne families. 
In all cases, these asteroids suffered impacts that could have stripped out the IDP external layer.
An exception are dark X-type asteroids such as 203 Pompeja, 409 Aspasia, and 212 Medea which do not belong to families and could have been considered intact planetesimals.
However, recent studies have shown that objects that were transferred into the main belt could have been already fragments themselves from the original planetesimal population 
\citep{Avdellidou2022,Avdellidou2024,Galinier2024}.
Note that \citet{Vernazza2025} proposed a similar ``onion-shell'' structure, in which chondritic porous IDPs constitute the surface component of trans-Neptunian objects (TNOs), and suggested that this model may also apply to large dark X- and D-type asteroids as well as TNOs.

\section{Conclusions}
We discovered two X-type objects exhibiting C-type spectral characteristics from observations at 0.45--2.5 \micron. 
Subsequent analysis reveals the following:

\begin{itemize}
\item 
In addition to the two X$\&$C asteroids we discovered, a survey of spectroscopic data for large asteroids in the literature revealed the existence of approximately ten X$\&$C type asteroids in total.
These X$\&$C asteroids exhibit a range of spectral properties between C-type and X-type asteroids, in terms of spectral slope and position of the hydration band.
The continuity between X-type and C-type spectra of asteroids suggests that these asteroids are (partially) differentiated bodies sharing the same primitive properties.
\item 
The absorption features at 1.1--1.3 \micron~in C-type asteroids are due to magnetite.
The reason for this lack of absorption in non-fresh CI1 chondrites is that alteration due to terrestrial weathering obscures the absorption of magnetite.
\item 
The surface of X$\&$C asteroids is more likely composed of cronstedtite than maghemite or very fine-grained magnetite.
X$\&$C asteroids may correspond to mantle layer of the differentiated CI chondrite parent bodies.
\end{itemize}

To further test this interpretation, it is important to obtain spectra of these asteroids at wavelengths longer than 2.5 \micron, similar to recent JWST observations of Jovian irregular satellites \citep{Sharkey2025}. 
We expect that future investigations of these objects over the 2.5--28 \micron~wavelength  range using \textit{JWST} \citep{Gardner2006}, \textit{SPHEREx} \citep{Crill2020}, and \textit{Ariel} \citep{Salvignol2024} will provide important constraints on their composition and origin.

\section*{Acknowledgements}
We thank the anonymous referee for the helpful suggestions.
We express our gratitude to Pierre Beck, J\'{e}r\^{o}me Gattacceca, and Adomas Valantinas for providing mineral and meteorite spectra. 
We are grateful to Takahiro Hiroi and Katsuhito Ohtsuka for their useful comments. 
We acknowledge Bin Yang for providing the spectroscopic data on the asteroids.
We acknowledge the sacred nature of MaunaKea and are grateful for the opportunity to conduct observations on this mountain.
Any opinions, findings, conclusions or recommendations expressed in this Article are those of the authors and do not necessarily reflect the views of NASA.
CT and FD are supported by the NASA grant 80NSSC18K0849 and 80NSSC22K0773.
IM acknowledges a grant from the Korean National Research Foundation (NRF) (MEST), funded by the Korean government (No.2023R1A2C1006180).
DK was supported by JSPS KAKENHI (grant nos. 23K03484).
MD is Leverhulme Visiting Professor at the University of Leicester with financial support from the Leverhulme Trust (UK).
UB thanks funding from an STFC PhD studentship and the LSST-DA Data Science Fellowship Program, which is funded by LSST-DA, the Brinson Foundation, the WoodNext Foundation, and the Research Corporation for Science Advancement Foundation; his participation in the program has benefited this work.
CA, UB, and MD were Visiting Astronomers at the Infrared Telescope Facility, which is operated by the University of Hawaii under contract 80HQTR19D0030 with the National Aeronautics and Space Administration.
This study was supported by JSPS KAKENHI (grant nos. 22H00179 and 24K00695) and by the Hypervelocity Impact Facility (former facility name: the Space Plasma Laboratory), ISAS, JAXA. 

\section*{Data Availability}
Data were mainly obtained from the Minor Planet Physical Properties Catalogue (https://mp3c.oca.eu). 
The \textit{Gaia} DR3 asteroid reflectance spectra were obtained from the archive website (https://archives.esac.esa.int/gaia). 
The new raw NIR spectra can be found in the NASA IRTF Data Archive when they become publicly available at The IRTF Data Archive website (https://irtfweb.ifa.hawaii.edu/research/irtf\_data\_archive.php). 
The reduced NIR spectra excluding those obtained from the `sp' run can be available upon request from the PI of the IRTF observing run (C. Avdellidou) or corresponding author.
The two `sp'  spectra executed as backups for the MITHNEOS MIT-Hawaii Near-Earth Object Spectroscopic Survey (http://smass.mit.edu/minus.html) are can be available upon request from Cristina A. Thomas or corresponding author.


\bibliographystyle{mnras}
\bibliography{XChase} 






\bsp	
\label{lastpage}
\end{document}